\documentstyle[12pt]{article}

\newcommand{\p}[1]{(\ref{#1})}
\begin{document}
\renewcommand{\thefootnote}{\fnsymbol{footnote}}
\thispagestyle{empty} {JINR E2-2000-193, \par CBPF-NF-029/00}

\vspace{.5cm}
\begin{center}
{\large\bf On the Classification of $N$-extended Supersymmetric\\
Quantum Mechanical Systems} \vspace{.5cm}\\

A. Pashnev ${}^a$\footnote{e-mail: pashnev@thsun1.jinr.ru}, and F.
Toppan ${}^b$\footnote{e-mail: toppan@cbpf.br} \vspace{.1cm}\\

${}^a${\it Bogoliubov Laboratory of Theoretical Physics, JINR} \\
{\it Dubna, 141980, Russia}\\ ${}^b${\it CBPF, DCP, Rua Dr. Xavier
Sigaud 150,\\ cep 22290-180 Rio de Janeiro (RJ),
Brazil}\vspace{.5cm}\\

{\bf Abstract}

\end{center}

In this paper some properties of the irreducible multiplets of
representation for the $N = ( p, q )$ -- extended supersymmetry in
one dimension are discussed. Essentially two results are here
presented. At first a peculiar property of the one dimension is
exhibited, namely that any multiplet containing $2d$ ($d$ bosonic
and $d$ fermionic) particles in $M$ different spin states, is
equivalent to a ${\bf {\bf \{d,d\}}}$ multiplet of just $2$ spin
states (all bosons and all fermions being grouped in the same
spin). Later, it is shown that the classification of all
multiplets of this kind carrying an irreducible representation of
the $N$ -- extended supersymmetry is in one-to-one correspondence
with the classification of real-valued Clifford $\Gamma$-matrices
of Weyl type. In particular, $p+q$ is mapped into $D$, the
space-time dimensionality, while $2d$ is determined to be the
dimensionality of the corresponding $\Gamma$-matrices. The
implications of these results to the theory of spinning particles
are analyzed.

\vfill \setcounter{page}0
\renewcommand{\thefootnote}{\arabic{footnote}}
\setcounter{footnote}0
\newpage
\section{Introduction}

Recently we have assisted at a regain of the interest in the
theory of Supersymmetric and Superconformal Quantum Mechanics due
to different physical motivations and viewpoints. Supersymmetric
and superconformal \cite{CDKKTP} (see also \cite{MS}) quantum
mechanical models succeed in describing the low-energy effective
dynamics, as well as the moduli space, of a certain class of black
holes. Another scenario involving Supersymmetric Quantum Mechanics
(SQM) concerns the light-cone quantization of supersymmetric
theories \cite{HP}. Besides that, (SQM)  models offer a natural
set-up for testing, under a rigorous mathematical framework, some
conjectures (like the $AdS/CFT$ correspondence for $AdS_2$) or
properties and consequences of dimensionally reduced
supersymmetric field theories \cite{CH} and such phenomena as
their spontaneous supersymmetry breaking
\cite{W}, \cite{W1}, including the partial breaking
 \cite{IKP}, \cite{DPRT}.
Having this in mind, the importance of the investigation of large
$N$-extended (SQM) models cannot be overestimated. Indeed, since
the reduced version to a one (temporal) dimension of a
supersymmetric $4d$ theory gets $4$ times the number of
supersymmetries of the original model, $N=2,4$ Super-Yang-Mills
are reduced to $N=8$ and respectively $N=16$ SQM models, while the
$N=8$ supergravity is associated with the $N=32$ SQM theory.
\par
Not
much attention has been paid however to such large--$N$ susy
quantum models and only partial results are known \cite{CR},
\cite{CH}. The reason however is clear, $N=4$ is the largest
number of extended supersymmetry for which a superfield formalism
is known.  Investigating the $N>4$ case requires
the use of component fields and computations soon become
cumbersome.
\par
 In
this paper we attack the problem of investigating large $N$
SQM models from a different
viewpoint. We are able to classify the irreducible multiplets of
representations of the $N$ extended supersymmetry. We prove at
first that all such multiplets are associated to fundamental short
multiplets in which all bosons and all fermions are accommodated
into just two spin states. In consequence of that, differently
presented dynamical systems turn out to be expression of the same
algebraic structure. Later, we give the full classification of the
short multiplets. We further mention how the above results find
application to the theory of the  particles with spin.
 \par
 The
closest references to the results here presented are given by the
papers \cite{{WTN},{GR}} in which the classification of (in our
language) the short multiplets for the Euclidean Supersymmetry was
derived.
\par
 In our work we further prove that all multiplets fit
into equivalence classes characterized by the short multiplets.
Besides that, we extend the classification of \cite{{WTN},{GR}} to
the pseudo-Euclidean Supersymmetry. Indeed, as  we will prove in
the following, it is in this larger class that symmetries of the
particles with spin moving in a Minkowskian or AdS-like background
should be looked for.

It is well known that the SQM,
being the simplest example of a theory which includes
simultaneously commuting and anticommuting variables, realizes as
its symmetry group the one -- dimensional supersymmetry. In
general this supersymmetry is generated by $N$ supercharges
$Q_i,\;\; i=1,2,\cdots,N$
 and the Hamiltonian
\begin{equation}\label{H}
H= - i \frac{\partial }{ \partial t}
\end{equation}
 with the
following algebra
\begin{equation}\label{Algebra}
\{ Q_i,Q_ j\}= {\omega}_{ij}H,
\end{equation}
where the constant tensor  ${\omega}_{ij}$  has $p$ positive and
$q$ negative eigenvalues. Usually all eigenvalues are positive and
the above algebra is named the $N$ -- extended one -- dimensional
supersymmetry.  Nevertheless, in general, reasons can exist
leading to an indefinite tensor ${\omega}_{ij}$ \cite{P}. In the
following, without loss of generality, the algebra of supercharges $Q_i$'s
will be conveniently diagonalized and normalized in such a way
that the tensor $\omega_{ij}$ can be expressed as
\begin{equation}\label{Diagonal}
\omega_{ij} =  \eta_{ij},
\label{metric}
\end{equation}
where $\eta_{ij}$ is a pseudo-Euclidean metric with the signature $(p,q)$.

The representation of the algebra \p{Algebra} is formed by
commuting (Bosonic) and anticommuting (Fermionic before the
quantization and Clifford after it) variables. Some of them are true
physical variables, others play an auxiliary role. Usually all
these variables are taken to be the components of irreducible
superfields.

The simplest way to construct a classical Lagrangian for the SQM
in $ D $ dimensions is to consider the superfields ($A=1,2,\cdots D$)
\begin{equation}\label{SF}
\Phi_A(\tau,\eta^\alpha)= \Phi_A^0(\tau)+
\eta^\alpha \Phi_{A\alpha}^1(\tau)+
\eta^{\alpha_1}\eta^{\alpha_2} \Phi_{A\alpha_1\alpha_2}^2(\tau)+\cdots+
\eta^{\alpha_1}\eta^{\alpha_2}\cdots\eta^{\alpha_N}
\Phi_{A\alpha_1\alpha_2\cdots \alpha_N}^N(\tau)
\end{equation}
in the superspace $(\tau,\eta^\alpha)$
with one bosonic coordinate
$\tau$ and $  N $
Grassmann coordinates $\eta^\alpha$.
%The series \p{SF} terminates due to the property of Grassmann coordinates
%$\eta^\alpha\eta^\alpha=0$ (no summation).
 Such superfields for general $N$ are
highly reducible and only lower values of $N$ were investigated
in details.
 The first components of the superfields are
the usual bosonic coordinates 
$ \Phi_A^0(\tau)$ , the next ones $\Phi_{A\alpha}^1(\tau)$
are the Grassmann coordinates. All the other components of the
superfields are auxiliary. So, the classical Lagrangian of the
SQM describes the evolution of
bosonic and additional Grassmann degrees of freedom, which after
quantization become generators of the Clifford algebra. This fact
naturally leads to the matrix realization of the Hamiltonian and
supercharges of SQM \cite{W}, \cite{C}, \cite{BM}.

The dimensionality of such realization depends on the total number
of Grassmann variables. In the case of scalar superfields \p{SF}
the dimensionality is $2^{[\frac{DN}{2}]}$. So, it rapidly growths
for extended supersymmetry.  The way out of this difficulty is to
use more complicated representations of the extended supersymmetry
\cite{AP} - \cite{BP2}. The simplest of them is given by the
chiral superfield, which contains one complex bosonic and
$\frac{N}{2} $ complex Grassmann fields. The Lagrangian for such
superfield naturally describes the two - dimensional SQM. The
ratio of numbers (fermi/boson) in this case is $\frac{N}{2} $
instead of $N$ as for scalar superfields. For more complicated
representations this ratio grows even more slowly \cite{IKL} -
\cite{BP2}. This fact has an essential influence on the
dimensionality of the matrix realizations of the Hamiltonian and
Supercharges - they are  smaller for the same number of bosonic
coordinates $\Phi_A^0$. The distinguishing feature of such
representations is the fact that the lowest components
$\Phi_A^0(\tau)$  ($A=1,2,\cdots D$) of the superfields \p{SF}
{\sl all together} form an irreducible representation of some
subgroup of the automorphism group of the algebra \p{Algebra}. The
corresponding actions are also  invariant under the
transformations of this subgroup which thus plays the role of
space (or space-time) rotations. In particular the inclusion of
the time coordinate $t(\tau)$ along with the space ones
$x^a(\tau)$ in an irreducible representation $\Phi_A^0(\tau)$ of
such subgroup means that such a subgroup, as well as the whole
automorphism group of the algebra \p{Algebra}, is pseudo-Euclidean
\cite{P}. In consequence of that the metric tensor $\eta_{ij}$ in
\p{Diagonal} is pseudo-Euclidean too.

So, the numbers of bosonic and fermionic physical components and,
correspondingly, the dimensionality of quantum Hamiltonian and
supercharges realized as matrices, crucially depend on the choice
of the irreducible superfield or, equivalently, irreducible
representation of the algebra \p{Algebra}. In this sense the
classification of all such representations  is very useful.

\setcounter{equation}0\section{The equivalence relations}

In this section we analyze the structure of the supermultiplets of
the $N$~--~extended supersymmetry in one dimensional space. We
will show in which sense all irreducible representations are
equivalent to the representations with a definite structure of the
multiplets.
 To fix the notation, let $N$
denotes the number of extended supersymmetries and $2d$ ($d$
bosons and $d$ fermions) the dimensionality of the corresponding
multiplet, (may be
reducible),
 carrying the $N$-extended SUSY representation.
In general such multiplet can be represented in the
form of a chain
\begin{equation}\label{rep}
\Phi^0_{a_0}, \quad \Phi^1_{a_1}, \quad \cdots ,\quad
\Phi^{M-1}_{a_{M-1}}, \quad \Phi^M_{a_M}
\end{equation}
whose components  $\Phi^I_{a_I}$, $
(a_I=1,2,\cdots,d_I)$ are real. All the components with even $I$ have the
same grassmann parity as that of $\Phi^0_{a_0}$, while the components with
odd values of $I$ have the opposite one.
 Such structure of the multiplet is
closely related with the superfield representations; the upper
index $I$, numbering the elements of the chain, corresponds to
their place in the superfield expansion \p{SF}   or to their
dimensionality which decreases by $1/2$ at each step along the
chain if one chooses
\begin{equation}\label{dim}
dim\; (\tau) =1, \quad dim\; (\eta) = \frac{1}{2}.
\end{equation}
 The number $M+1$  is the
length of the supermultiplet, $M$ being subjected to the
constraint $M\leq N$ since, for example in the case of irreducible
representations, not all the components of the superfield \p{SF}
are independent - some of the higher components are expressed in
terms of time derivatives of the lower ones.

For the supermultiplet \p{rep} we will also use the short
notation ${\bf \{d_0, d_1, \cdots ,d_{M}\}}$.
 As an example, in the case of $N=2$ one can
consider the irreducible representations ${\bf \{1, 2, 1\}}$ (real
superfield)
\begin{equation}
\label{113}
\Phi=\Phi(\tau, \eta_1, \eta_2)=
\Phi^0(\tau)+i\eta^\alpha\Phi^1_\alpha(\tau)
+i\eta^1\eta^2 \Phi^2(\tau),
\end{equation}
and ${\bf \{2, 2\}}$ (chiral superfield)
\begin{equation}
\label{114}
\tilde\Phi(\tau,\eta,\overline{\eta})=
\tilde\Phi^0(\tau)+\eta\tilde\Phi^1(\tau)+
\frac{i}{2}\overline{\eta}\eta\dot{\tilde\Phi}^0(\tau),
\end{equation}
where $\eta=\eta_1+i\eta_2,\quad
\overline{\eta}=\eta_1-i\eta_2\quad$ are complex Grassmann
coordinates. The last component in the expression \p{114} is
proportional to the time derivative of the first one. Both
$\tilde\Phi^0(\tau)$ and $\tilde\Phi^1(\tau)$ in \p{114} are
complex. Obviously, in both cases \p{113} and \p{114}, $\sum_I
d_I=2$ separately for real bosonic and fermionic components.

Due to a dimensionality arguments the supersymmetry transformation law for the
components $\Phi^I_{a_I}$ is of the following form
($\varepsilon^i$ are infinitesimal Grassmann parameters)
\begin{equation}\label{genlaw}
  \delta_\varepsilon \Phi^I_{a_I} = \varepsilon^i
{(C^I_i)_{a_I}}^{a_{I+1}}\Phi^{I+1}_{a_{I+1}} + \varepsilon^i
{(\tilde{C}^I_i)_{a_I}}^{a_{I-1}}\frac{d}{d\tau}{\Phi}^{I-1}_{a_{I-1}}.
\end{equation}
Evidently, due to the absence in
\p{rep} of the components with $I=-1,\quad I=M+1$, the
transformation laws for the end components of the chain are
simpler:
\begin{equation}\label{endlaw}
  \delta_\varepsilon \Phi^0_{a_0} = \varepsilon^i
{(C^0_i)_{a_0}}^{a_{1}}\Phi^{1}_{a_{1}}, \quad
\delta_\varepsilon \Phi^M_{a_M} =
  \varepsilon^i
{(\tilde{C}^M_i)_{a_M}}^{a_{M-1}}\frac{d}{d\tau}{\Phi}^{M-1}_{a_{M-1}}.
\end{equation}
The transformation law for the last component of the multiplet
reads that it transforms as a total derivative. It is a very
essential property, because the integral of this component is
invariant under the supersymmetry transformations and can be used
to construct invariant actions.

Another very important consequence of the transformation law of
the last components is present only in one dimension. Just in this
case one can redefine this component
\begin{equation}\label{redef}
  \Phi^M_{a_M}=\frac{d}{d\tau}\Psi^{M-2}_{a_M}
\end{equation}
in terms of some functions $\Psi^{M-2}_{a_M}$. This correspondence
is exact up to some constants $C^M_{a_M}$ which describe the
trivial representations of the supersymmetry algebra. The
dimensionality of the new components $\Psi^{M-2}_{a_M}$ coincides
with the dimensionality of the components $\Phi^{M-2}_{a_{M-2}}$.
Moreover, their transformation law is of the same type -- they
transform through the components ${\Phi}^{M-1}_{a_{M-1}}$ and
${\Phi}^{M-3}_{a_{M-3}}$ (with vanishing coefficients before the
time derivative of the last ones)
\begin{equation}\label{last}
 \delta_\varepsilon \Psi^{M-2}_{a_M} =
  \varepsilon^i
{(\tilde{C}^M_i)_{a_M}}^{a_{M-1}}{\Phi}^{M-1}_{a_{M-1}}.
\end{equation}
So, we have shown that up to trivial representations of the
supersymmetry algebra the supermultiplet ${\bf \{d_0, d_1, \cdots
,d_{M-2}, d_{M-1}, d_{M}\}}$ is equivalent to the supermultiplet\\
${\bf \{d_0, d_1, \cdots, d_{M-2}+d_M, d_{M-1}, 0\}}$. It means
that the initial supermultiplet of length $M+1$ is equivalent to a
shorter multiplet of length $M$. Evidently, the total number of
bosonic and fermionic components in both supermultiplets is the
same. This procedure can obviously be repeated $M-1$ times, so
that at the end one reaches the shortest multiplet of length $2$ -
the multiplet ${\bf \{d, d\}}$.

The simplest example of such shortening of the length is given in
the case of $N=2$. The component $\Phi^2(\tau)$ in the real
superfield \p{113} transforms as a total derivative of some new
field  $\Psi^0(\tau)$ which, together with  $\Phi^0(\tau)$, forms
the complex $\tilde\Phi^0(\tau)$ of \p{114}. A more complicated
example is furnished by the $N=4$ representation multiplets ${\bf
\{1,4,3\}}$, ${\bf \{2,4,2\}}$ and ${\bf \{3,4,1\}}$ which were
used in \cite{{IKL},{BP1},{IKP}} for one-dimensional, \cite{BP2}
for two-dimensional and \cite{{IS},{BP}} for three-dimensional
SQM respectively. The corresponding
components of these representations are interconnected by the
transformation \p{redef}. To our knowledge, the multiplet $(4,4)$
which should be useful in four-dimensional case was not considered
in the literature.

In principle, one can consider the inverse
procedure - the lengthening of the multiplet starting from the
${\bf \{d,d\}}$'s one. Only the first step is trivial - the
transition from the ${\bf \{d,d\}}$ multiplet to the ${\bf
\{d-d_1,d,d_1\}}$ multiplet can always be done with the help of
the transformation inverse to \p{redef} applied to an arbitrary
number $d_1\leq d$ of the first components of the initial
multiplet. The possibility of further lengthening must be analyzed
separately in each particular case and will be shortly discussed
at the end of the paper.
%is out of the scope of the present paper.
\par
It should be noticed that $d_1= d$ is allowed.
The corresponding transformation links two length-$2$ multiplets,
the one in which bosons have higher spin, to the one in which
fermions have higher spins. Explicitly we have
\begin{equation}\label{twolaw}
  \delta_\varepsilon \Phi_{a} = \varepsilon^i
{(C_i)_{a}}^{b}\Psi_{b}, \quad \delta_\varepsilon \Psi_a =
\varepsilon^i {(\tilde{C}_i)}_{a}^{b}\frac{d}{d\tau}{\Phi}_{b},
\end{equation}
while for $\Xi_a=\frac{d}{d\tau}{\Psi}_{a}$ we get
\begin{equation}\label{newtwolaw}
  \delta_\varepsilon \Xi_{a} = \varepsilon^i
{(C_i)_{a}}^{b}\Phi_{b}, \quad \delta_\varepsilon \Phi_a =
\varepsilon^i {(\tilde{C}_i)}_{a}^{b}\frac{d}{d\tau}{\Xi}_{b}.
\end{equation}
We finally comment that, due to the previous considerations, the
classification of all supermultiplets of length $2$ automatically
provides the classification of all supermultiplets of length $3$.
In many physical application of interest, this is quite
sufficient.

\setcounter{equation}0\section{Extended supersymmetries and
real-valued\\ Clifford algebras}

The main result of the previous Section is that the problem of
classifying all $N$-extended supersymmetric quantum mechanical
systems is reduced to
 the problem of classifying the irreducible
representations \p{rep} of length $2$. Having this in mind we
simplify the notations. Let the indices
 $a, \alpha = 1,
\cdots, d$ number the bosonic (and respectively fermionic)
elements in the SUSY multiplet. All  of them are assumed to depend on
the time coordinate $\tau$ ($X_a\equiv X_a(\tau)$,
$\theta_\alpha\equiv \theta_\alpha (\tau)$).
\par
In order to be definite and without loss of generality let us take
the bosonic elements to be the first ones in the chain ${\bf \{d,d\}}$,
which can be conveniently represented also as a column
\begin{equation}\label{basis}
\Psi=\left(
\begin{array}{c}
X_a \\
\theta_\alpha
\end{array}
\right),
\end{equation}
the equations \p{genlaw} are reduced
to the following set of equations
\begin{eqnarray}\label{genlaw1}
\delta_\varepsilon X_a &=& \varepsilon^i
{(C_i)_a}^{\alpha}\theta_{\alpha}\equiv i(\varepsilon^iQ_i \Psi)_a
\nonumber\\
\delta_\varepsilon
\theta_{\alpha} &=&\varepsilon^i
{(\tilde{C}_i)_{\alpha}}^b\frac{d}{d\tau} X_b\equiv i(\varepsilon^iQ_i
\Psi)_\alpha
\label{transf}
\end{eqnarray}
where, as a consequence of \p{Algebra},
\begin{eqnarray}
C_i {\tilde C}_j + C_j{\tilde C}_i &=& i \eta_{ij}
\end{eqnarray}
and
\begin{eqnarray}
{\tilde C}_i C_j + {\tilde C}_j C_i &=& i\eta_{ij}
\end{eqnarray}

Since $\varepsilon_i, X_a, \theta_\alpha$ are  real, the matrices
 $C_i$'s, ${\tilde C}_i$'s have to be respectively
imaginary and real.
 If we set (just for normalization)
\begin{eqnarray}
C_i &=&  \frac{i}{\sqrt{2}} \sigma_i\nonumber\\\label{sign}
 {\tilde C}_i&=&\frac{1}{\sqrt{2}}{\tilde\sigma}_i
\end{eqnarray}
and accommodate $ \sigma_i, {\tilde\sigma}_i$ into a single matrix
\begin{equation}\label{Gamma}
\Gamma_i=\left(
\begin{array}{cc}
0 & \sigma_i \\
{\tilde\sigma}_i& 0
\end{array}
\right),
\end{equation}
they form a set of real-valued Clifford $\Gamma$-matrices of Weyl
type (i.e. block antidiagonal), obeying the (pseudo-) Euclidean
anticommutation relations
\begin{eqnarray}\label{GG}
\{\Gamma_i, \Gamma_j \} &=& 2 \eta_{ij}.
\end{eqnarray}
Conversely, given a set of (pseudo-~) Euclidean real-valued
Clifford $\Gamma$-matrices of Weyl type, one can invert the above
procedure and reconstruct the supercharges $Q_i$
\begin{equation}
Q_i=\frac{1}{\sqrt{2}}\left(
\begin{array}{cc}
0 & \sigma_i \\
{\tilde\sigma}_i\cdot H& 0
\end{array}
\right)
\end{equation}
in the basis \p{basis}.

In addition to the matrices $\Gamma^i$ \p{Gamma} in the space of
vectors \p{basis} the further matrix $\Gamma^{N+1}$, which
anticommutes with the supercharges and corresponds to the
fermionic number, exists
\begin{equation}\label{Gamma5}
\Gamma_{N+1}=\left(
\begin{array}{cc}
1 & 0 \\
0& -1
\end{array}
\right).
\end{equation}

Altogether the matrices \p{Gamma} and \p{Gamma5} form the
real-valued representation $\Gamma_I$ of the (pseudo-~) Euclidean
Clifford algebra with the signature $(p+1,q)$.

Instead of \p{sign} one can take
\begin{eqnarray}
C_i &=&  \frac{i}{\sqrt{2}} \sigma_i\nonumber\\\label{sign1}
 {\tilde C}_i&=&-\frac{1}{\sqrt{2}}{\tilde\sigma}_i
\end{eqnarray}
and accommodate $ \sigma_i, {\tilde\sigma}_i$ into the matrices \p{Gamma}
which now
obey the (pseudo-) Euclidean
anticommutation relations
\begin{eqnarray}\label{GG1}
\{\tilde\Gamma_i, \tilde\Gamma_j \} &=& -2 \eta_{ij}
\end{eqnarray}
with opposite to \p{GG} sign of the righthand side.
Together with fermion
number matrix \p{Gamma5} new matrix $\tilde\Gamma_i$ form the
real-valued representation of the (pseudo-~) Euclidean
Clifford algebra with the signature $(q+1,p)$. This fact means that the
representations of $C_{p+1,q}$ and $C_{q+1,p}$ should be connected one with
the other. Indeed, this connection is established by the correspondence
\begin{equation}\label{equiv}
  \tilde\Gamma_i=\Gamma_{N+1}\Gamma_i.
\end{equation}

Thus, the
representations of the $(p,q)$- extended supersymmetry algebra
\p{Algebra} are in one-to-one correspondence with the real-valued
representations of the Clifford algebra $C_{p+1,q} \sim C_{q+1,p}$.

In general the real Clifford algebras were classified in
\cite{ABS} (for the compact case $q=0$) and in \cite{Porteus} (for
the noncompact case). The construction along the lines
\p{genlaw1}-\p{Gamma5} for representations of the type ${\bf \{d,d\}}$
in the case of positively definite signature $(p,q) = (N,0)$ was
performed in \cite{WTN} (see also \cite{GR}) where the
dimensionalities as well as realizations of the $\Gamma$-matrices
\p{Gamma} were described. In the case of pseudo-Euclidean metric
with signature $(p, q)$ such construction extensively uses the
considerations of the  papers \cite{O}. The results will be
presented in the next section.

\setcounter{equation}0\section{Classification of the irreducible
representations.}

According to the previous Section results, the classification of
irreducible multiplets of representation of a $(p,q)$ extended
supersymmetry is in one -- to -- one correspondence with the
classification of the real Clifford algebras $C_{p,q}$ with the
further property that the $\Gamma$ matrices can be realized in
Weyl (i.e. block antidiagonal) form. \par For what concerns real
matrix representations of the Clifford algebras we borrow the
results of \cite{O}. Three cases have to be distinguished for real
representations, specified by the type of most general solution
allowed for a real matrix $S$ commuting with all the Clifford
$\Gamma_i$ matrices, i.e.\\
 {\em i)} the normal case, realized
when $S$ is a multiple of the identity, \\
{\em ii)} the almost
complex case, for $S$ being given by a linear combination of the
identity and of a real $J^2= -{\bf 1}$ matrix,\\
{\em iii)} finally
the quaternionic case, for $S$ being a linear combination of real
matrices satisfying the quaternionic algebra.
\par
Real irreducible
representations of normal type exist whenever the condition\\ $p-q
= 0,1,2 \quad mod \quad 8$ is satisfied (their dimensionality
being given by $2^{[\frac{N}{2}]}$, where $N=p+q$), while the
almost complex and the quaternionic type representations
 are realized in the $p-q
= 3,7 \quad mod \quad 8$ and in the $p-q = 4,5,6 \quad mod \quad
8$ cases respectively. The dimensionality of these representations
is given in both cases by $2^{[\frac{N}{2}]+1}$.\par We further
require the extra-condition that the real representations should
admit a block antidiagonal realization for the Clifford $\Gamma$
matrices. This condition is met for $p-q = 0 \quad mod \quad 8$ in
the normal case (it corresponds to the standard Majorana-Weyl
requirement), $p-q = 7 \quad mod \quad 8$ in the almost complex
case and  $p-q = 4,6 \quad mod \quad 8$ in the quaternionic case.
In all these cases the real irreducible representation is unique.
\par The above results can be summarized as follows,
expressing the dimensionality of the irreducible representations
of the algebra \p{Algebra} (independently of the length $M+1$ of
the chain \p{rep}) as function of the signature $(p,q)$. Let
$q=8k+m,\quad 0\leq m \leq 7$ and $p=8l+m+n,\quad 1\leq n \leq 8$
($l=-1$ when $k=0$ and $p\leq q$ ). Then, the dimensionalities of
the bosonic (fermionic) spaces are given by the expression
\begin{equation}\label{dimensions}
  d=2^{4k+4l+m}\cdot G(n),
\end{equation}
where the so called Radon-Hurwitz function $G(n)$ is defined with
the help of the table \cite{ABS}
\begin{equation}\label{G}
\begin{tabular}{c|c|c|c|c|c|c|c|c}
  $ n $  & 1 & 2 & 3 & 4 & 5 & 6 & 7 & 8 \\ \hline
  $ G(n)$ & 1 & 2 & 4 & 4 & 8 & 8 & 8 & 8 \\
 \end{tabular}
\end{equation}
By words, $G(n)=2^r$, where $r$ is the nearest integer which is
greater or equal to $\log_{2}n$.

A second useful table expresses conversely which kind of
signatures $(p,q)$ are possible for a given dimensionality of the
bosonic and fermionic spaces. In order to do so it is convenient
to introduce the notion of maximally extended supersymmetry. The
$C_{p,q}$ ($p-q= 6\quad mod \quad 8$) real representation for the
quaternionic case can be recovered from the $7 \quad mod \quad 8$
almost complex $C_{p+1,q}$ representation by deleting one of the
$\Gamma$ matrices; in its turn the latter representation is
recovered from the $C_{p+2,q}$ normal Majorana-Weyl representation
by deleting another $\Gamma$ matrix. The dimensionality of the
three representations above being the same, the normal
Majorana-Weyl representation realizes the maximal possible
extension of supersymmetry compatible with the dimensionality of
the representation. In search for the maximal extension of
supersymmetry we can therefore limit ourselves to consider the
normal Majorana-Weyl representations, as well as the quaternionic
ones satisfying the $p-q = 4\quad mod \quad 8$ condition.\par Let
therefore be $p=8l+m+8+4\epsilon$ and $q= 8 k + m$, where the
range of values for $k,l,m$ is the same as before, while
$\epsilon$ assumes two values, distinguishing the Majorana-Weyl
($\epsilon =0 $) and the quaternionic case ($\epsilon = 1$). A
space of $d=2^t$ bosonic and $d= 2^t$ fermionic states can carry
the following set of maximally extended supersymmetries
\begin{equation}\label{Inverse1}
(p= t-4z + 5 - 3\epsilon, q= t+4z+\epsilon - 3)
\end{equation}
where the integer $z= k-l$ must take values in the interval
\begin{equation}
\frac{1}{4}(3-t-\epsilon ) \leq z \leq \frac{1}{4} (t+ 5-
3\epsilon )
\end{equation} in order to guarantee the $p\geq 0$
and $q\geq 0 $ requirements.
It is convenient also to represent the answer by the following table
\begin{center}
\begin{equation}\label{Inverse}
\begin{tabular}{c|c}
 $ d $  & $(p, q)$ \\\hline
 $\phantom{+}$&$\phantom{+}$\\
  $2^{4l}$ & $(8l-4k+1,  4k+1), (8l-4k-2,  4k+2)$ \\
 $\phantom{+}$&$\phantom{+}$\\
  $2^{4l+1}$ & $(8l-4k+2,  4k+2),  (8l-4k-1,  4k+3)$ \\
 $\phantom{+}$&$\phantom{+}$\\
 $2^{4l+2}$  & $(8l-4k+4,  4k),  (8l-4k+3,  4k+3)$ \\
 $\phantom{+}$&$\phantom{+}$\\
$2^{4l+3}$ & $(8l-4k+8,  4k),  (8l-4k+5,  4k+1)$ \\
  \end{tabular}
\end{equation}
\end{center}
where $k$ is integer satisfying the only conditions $p\geq 0,\quad q\geq 0$.
\par
For the lowest values of dimensionality $d$ the solutions are given
by the table:
\begin{center}
\begin{equation}\label{Low}
\begin{tabular}{c|c}
 $ d $  & $(p, q)$ \\\hline
 $\phantom{+}$&$\phantom{+}$\\
 $1$& $(1, 1)$ \\
 $\phantom{+}$&$\phantom{+}$\\
  $2$ & $(2, 2)$ \\
 $\phantom{+}$&$\phantom{+}$\\
 $4$  & $(4, 0), (3, 3), (0, 4)$ \\
 $\phantom{+}$&$\phantom{+}$\\
$8$ & $(8, 0), (5, 1), (4, 4), (1, 5), (0, 8)$ \\
$\phantom{+}$&$\phantom{+}$\\
$16$ & $(9, 1), (6, 2), (5, 5), (2, 6), (1, 9)$ \\
$\phantom{+}$&$\phantom{+}$\\
$32$ & $(10, 2), (7, 3), (6, 6), (3, 7), (2, 10)$ \\
  \end{tabular}
\end{equation}
\end{center}

As already recalled, obviously the representations $(p',q')$ with
$p'\leq p, q'\leq q$ also exist for the same dimensionality $d$.
These representations are also irreducible unless either $p'$ or
$q'$ become too small. For example, the $d=16$-dimensional
representations are irreducible not only for the signature
$(p,q)=(5,5)$, but also for the pairs $(5,4), (5,3), (5,2), (4,5),
(3,5), (2,5)$, while the irreducible representations for the
signatures $(5, 1), (4, 4), (1, 5)$ are encountered in $d=8$
dimensions.

\setcounter{equation}0\section{Examples of representations for supercharges.}
 For the case $(p,q)=(4,0)$ the following matrices realize
four supercharges: {\scriptsize
\begin{eqnarray}\nonumber
\begin{tabular}{l|llll|llll|}
             &$0$&$0$&$0$&$0$&$0$&$0$&$0$&$1$\\
             &$0$&$0$&$0$&$0$&$0$&$0$&$1$&$0$\\
             &$0$&$0$&$0$&$0$&$0$&$1$&$0$&$0$\\
$Q_1=\frac{1}{\sqrt{2}}$&$0$&$0$&$0$&$0$&$1$&$0$&$0$&$0$\\\cline{2-9}
             &$0$&$0$&$0$&$H$&$0$&$0$&$0$&$0$\\
             &$0$&$0$&$H$&$0$&$0$&$0$&$0$&$0$\\
             &$0$&$H$&$0$&$0$&$0$&$0$&$0$&$0$\\
             &$H$&$0$&$0$&$0$&$0$&$0$&$0$&$0$\\
\end{tabular}
&&
\begin{tabular}{c|cccc|cccc|}
             &$0$&$0$&$0$&$0$&$0$&$0$&$1$&$0$\\
             &$0$&$0$&$0$&$0$&$0$&$0$&$0$&$-1$\\
             &$0$&$0$&$0$&$0$&$1$&$0$&$0$&$0$\\
$Q_2=\frac{1}{\sqrt{2}}$&$0$&$0$&$0$&$0$&$0$&$-1$&$0$&$0$\\\cline{2-9}
             &$0$&$0$&$H$&$0$&$0$&$0$&$0$&$0$\\
             &$0$&$0$&$0$&$-H$&$0$&$0$&$0$&$0$\\
             &$H$&$0$&$0$&$0$&$0$&$0$&$0$&$0$\\
             &$0$&$-H$&$0$&$0$&$0$&$0$&$0$&$0$ \\
\end{tabular}\\
\label{example}
&&  \\\nonumber
\begin{tabular}{c|cccc|cccc|}
             &$0$&$0$&$0$&$0$&$1$&$0$&$0$&$0$\\
             &$0$&$0$&$0$&$0$&$0$&$1$&$0$&$0$\\
             &$0$&$0$&$0$&$0$&$0$&$0$&$-1$&$0$\\
$Q_3=\frac{1}{\sqrt{2}}$&$0$&$0$&$0$&$0$&$0$&$0$&$0$&$-1$\\\cline{2-9}
             &$H$&$0$&$0$&$0$&$0$&$0$&$0$&$0$\\
             &$0$&$H$&$0$&$0$&$0$&$0$&$0$&$0$\\
             &$0$&$0$&$-H$&$0$&$0$&$0$&$0$&$0$\\
             &$0$&$0$&$0$&$-H$&$0$&$0$&$0$&$0$\\
\end{tabular}
&&
\begin{tabular}{c|cccc|cccc|}
             &$0$&$0$&$0$&$0$&$0$&$1$&$0$&$0$\\
             &$0$&$0$&$0$&$0$&$-1$&$0$&$0$&$0$\\
             &$0$&$0$&$0$&$0$&$0$&$0$&$0$&$-1$\\
$Q_4=\frac{1}{\sqrt{2}}$&$0$&$0$&$0$&$0$&$0$&$0$&$1$&$0$\\\cline{2-9}
             &$0$&$-H$&$0$&$0$&$0$&$0$&$0$&$0$\\
             &$H$&$0$&$0$&$0$&$0$&$0$&$0$&$0$\\
             &$0$&$0$&$0$&$H$&$0$&$0$&$0$&$0$\\
             &$0$&$0$&$-H$&$0$&$0$&$0$&$0$&$0$ \\
\end{tabular}
\end{eqnarray}
}
These supercharges act in the space with $4$ bosonic and $4$
fermionic coordinates forming the representation ${\bf \{4,4\}}$. The
automorphism group
 $SO(p,q)$ of the algebra \p{Algebra}  is now $SO(4)$.

Besides the transformations of the automorphism group $Q_i'=\Lambda_i^jQ_j$
the algebra of supercharges is
invariant under the more general transformations
of the type
\begin{equation}\label{newtrans}
  Q_i'= U Q_i U^{-1}
\end{equation}
with  block-diagonal $8\times 8$ matrices $U$. When the matrix $U$
is nonsingular and real the transformation \p{newtrans} simply
means a change of basis in bosonic and fermionic sectors. {}On the
other hand this transformation drastically changes the
representation when $U$ depends on the operator $H=-id/d\tau$. In
this case the transformation \p{newtrans} is in general nonlocal.
Nevertheless, transformations exist which do not lead to any
nonlocality. In particular, in the framework of the example
\p{example} one can take
\begin{equation}\label{U1}
  U_1=diag\{1,1,1,H,1,1,1,1\}
\end{equation}
and obtain the new realization for the operators $Q_i$
{\scriptsize
\begin{eqnarray}\nonumber
\begin{tabular}{l|llll|llll|}
             &$0$&$0$&$0$&$0$&$0$&$0$&$0$&$1$\\
             &$0$&$0$&$0$&$0$&$0$&$0$&$1$&$0$\\
             &$0$&$0$&$0$&$0$&$0$&$1$&$0$&$0$\\
$Q_1=\frac{1}{\sqrt{2}}$&$0$&$0$&$0$&$0$&$H$&$0$&$0$&$0$\\\cline{2-9}
             &$0$&$0$&$0$&$1$&$0$&$0$&$0$&$0$\\
             &$0$&$0$&$H$&$0$&$0$&$0$&$0$&$0$\\
             &$0$&$H$&$0$&$0$&$0$&$0$&$0$&$0$\\
             &$H$&$0$&$0$&$0$&$0$&$0$&$0$&$0$\\
\end{tabular}
&&
\begin{tabular}{c|cccc|cccc|}
             &$0$&$0$&$0$&$0$&$0$&$0$&$1$&$0$\\
             &$0$&$0$&$0$&$0$&$0$&$0$&$0$&$-1$\\
             &$0$&$0$&$0$&$0$&$1$&$0$&$0$&$0$\\
$Q_2=\frac{1}{\sqrt{2}}$&$0$&$0$&$0$&$0$&$0$&$-H$&$0$&$0$\\\cline{2-9}
             &$0$&$0$&$H$&$0$&$0$&$0$&$0$&$0$\\
             &$0$&$0$&$0$&$-1$&$0$&$0$&$0$&$0$\\
             &$H$&$0$&$0$&$0$&$0$&$0$&$0$&$0$\\
             &$0$&$-H$&$0$&$0$&$0$&$0$&$0$&$0$ \\
\end{tabular}\\\label{example1}
&&  \\\nonumber
\begin{tabular}{c|cccc|cccc|}
             &$0$&$0$&$0$&$0$&$1$&$0$&$0$&$0$\\
             &$0$&$0$&$0$&$0$&$0$&$1$&$0$&$0$\\
             &$0$&$0$&$0$&$0$&$0$&$0$&$-1$&$0$\\
$Q_3=\frac{1}{\sqrt{2}}$&$0$&$0$&$0$&$0$&$0$&$0$&$0$&$-H$\\\cline{2-9}
             &$H$&$0$&$0$&$0$&$0$&$0$&$0$&$0$\\
             &$0$&$H$&$0$&$0$&$0$&$0$&$0$&$0$\\
             &$0$&$0$&$-H$&$0$&$0$&$0$&$0$&$0$\\
             &$0$&$0$&$0$&$-1$&$0$&$0$&$0$&$0$\\
\end{tabular}
&&
\begin{tabular}{c|cccc|cccc|}
             &$0$&$0$&$0$&$0$&$0$&$1$&$0$&$0$\\
             &$0$&$0$&$0$&$0$&$-1$&$0$&$0$&$0$\\
             &$0$&$0$&$0$&$0$&$0$&$0$&$0$&$-1$\\
$Q_4=\frac{1}{\sqrt{2}}$&$0$&$0$&$0$&$0$&$0$&$0$&$H$&$0$\\\cline{2-9}
             &$0$&$-H$&$0$&$0$&$0$&$0$&$0$&$0$\\
             &$H$&$0$&$0$&$0$&$0$&$0$&$0$&$0$\\
             &$0$&$0$&$0$&$1$&$0$&$0$&$0$&$0$\\
             &$0$&$0$&$-H$&$0$&$0$&$0$&$0$&$0$ \\
\end{tabular}
\end{eqnarray}
}
in which all the elements of the last column in the left
off-diagonal block have lost the multiplier $H$. Instead, all the
elements of the last arrow in the right off-diagonal block
acquired $H$ as a multiplier. This representation of the
supercharges corresponds to the irreducible supermultiplet ${\bf
\{3,4,1\}}$ which was used in \cite{{IS},{BP}} for constructing
the $3$-dimensional $N=4$ extended SQM. The supermultiplets ${\bf
\{2,4,2\}}$ and ${\bf \{1,4,3\}}$ are derived with the help of the
following matrices $U$
\begin{equation}\label{U23}
  U_2=diag\{1,1,H,H,1,1,1,1\},\;
U_3=diag\{1,H,H,H,1,1,1,1\}.
\end{equation}
The next one in this sequence
\begin{equation}\label{U4}
U_4=diag\{H,H,H,H,1,1,1,1\}
\end{equation}
gives again the supermultiplet ${\bf \{4,4\}}$ but with the opposite
grading - the first in the chain is the fermionic subspace. This
completes the classification of the irreducible supermultiplets of
the $N=4$ extended SQM. One can show that all the irreducible
supermultiplets of the $(p,q)$ extended SQM are of length which
does not exceed $3$ when the constraint
\begin{equation}\label{length3}
  d\leq p+q
\end{equation}
is fulfilled. The determination of the possible values of the
lengths of irreducible supermultiplets, as well as their detailed
structure, in the case when \p{length3} is not fulfilled needs a
separate investigation.
\par
A simple example of an irreducible supermultiplet of length 4 is
given by the $(p,q)=(3,0)$ case, in which the irreducible
representation has also $d=4$ and supercharges in the ${\bf
\{4,4\}}$ representation are given by $Q_1, Q_2, Q_3$ in
\p{example}. Taking
\begin{equation}\label{U5}
U_5=diag\{1,H,H,H,1,H,1,1\}
\end{equation}
one derives the expressions for all $4$ supercharges {\scriptsize
\begin{eqnarray}\nonumber
Q_1&\!\!\!\!\!\!\!\!\!\!=&\!\!\!\!\!\!\!\!\!\!\frac{1}{\sqrt{2}}
\begin{tabular}{|llll|llll|}
             $0$&$0$&$0$&$0$&$0$&$0$&$0$&$1$\\
             $0$&$0$&$0$&$0$&$0$&$0$&$H$&$0$\\
             $0$&$0$&$0$&$0$&$0$&$1$&$0$&$0$\\
             $0$&$0$&$0$&$0$&$H$&$0$&$0$&$0$\\\cline{1-8}
             $0$&$0$&$0$&$1$&$0$&$0$&$0$&$0$\\
             $0$&$0$&$H$&$0$&$0$&$0$&$0$&$0$\\
             $0$&$1$&$0$&$0$&$0$&$0$&$0$&$0$\\
             $H$&$0$&$0$&$0$&$0$&$0$&$0$&$0$\\
\end{tabular}
\quad\quad\quad\;\;
Q_2=\frac{1}{\sqrt{2}}
\begin{tabular}{|cccc|cccc|}
             $0$&$0$&$0$&$0$&$0$&$0$&$1$&$0$\\
             $0$&$0$&$0$&$0$&$0$&$0$&$0$&$-H$\\
             $0$&$0$&$0$&$0$&$H$&$0$&$0$&$0$\\
             $0$&$0$&$0$&$0$&$0$&$-1$&$0$&$0$\\\cline{1-8}
             $0$&$0$&$1$&$0$&$0$&$0$&$0$&$0$\\
             $0$&$0$&$0$&$-H$&$0$&$0$&$0$&$0$\\
             $H$&$0$&$0$&$0$&$0$&$0$&$0$&$0$\\
             $0$&$-1$&$0$&$0$&$0$&$0$&$0$&$0$ \\
\end{tabular}\\
\label{example2}
&&  \\\nonumber
Q_3&\!\!\!\!\!\!\!\!\!\!=&\!\!\!\!\!\!\!\!\!\!\frac{1}{\sqrt{2}}
\begin{tabular}{|cccc|cccc|}
             $0$&$0$&$0$&$0$&$1$&$0$&$0$&$0$\\
             $0$&$0$&$0$&$0$&$0$&$1$&$0$&$0$\\
             $0$&$0$&$0$&$0$&$0$&$0$&$-H$&$0$\\
             $0$&$0$&$0$&$0$&$0$&$0$&$0$&$-H$\\\cline{1-8}
             $H$&$0$&$0$&$0$&$0$&$0$&$0$&$0$\\
             $0$&$H$&$0$&$0$&$0$&$0$&$0$&$0$\\
             $0$&$0$&$-1$&$0$&$0$&$0$&$0$&$0$\\
             $0$&$0$&$0$&$-1$&$0$&$0$&$0$&$0$\\
\end{tabular}
\;\;
Q_4=\frac{1}{\sqrt{2}}
\begin{tabular}{|cccc|cccc|}
             $0$&$0$&$0$&$0$&$0$&$H^{-1}$&$0$&$0$\\
             $0$&$0$&$0$&$0$&$-H$&$0$&$0$&$0$\\
             $0$&$0$&$0$&$0$&$0$&$0$&$0$&$-H$\\
             $0$&$0$&$0$&$0$&$0$&$0$&$H$&$0$\\\cline{1-8}
             $0$&$-1$&$0$&$0$&$0$&$0$&$0$&$0$\\
             $H^2$&$0$&$0$&$0$&$0$&$0$&$0$&$0$\\
             $0$&$0$&$0$&$1$&$0$&$0$&$0$&$0$\\
             $0$&$0$&$-1$&$0$&$0$&$0$&$0$&$0$ \\
\end{tabular}
\end{eqnarray}}
\noindent from which one can easily see that the fourth
supercharge $Q_4$ becomes singular after the transformation.
Indeed, just the first three supercharges in \p{example2} are
realized in the irreducible representation ${\bf \{1,3,3,1\}}$ of
length $4$.

 The Weyl-type $C_{0,4}$ representation
has been explicitly presented in \cite{O}. Due to the $mod \quad
8$ property of $\Gamma$ matrices, it allows, together with
$C_{4,0}$, to construct all quaternionic representations of Weyl
type for the allowed values of $(p,q)$. For what concerns the
Majorana-Weyl representations, an algorithm to explicitly
construct them can be found e.g. in \cite{DRT}. Moreover,
the following symmetry property
\begin{equation}\label{Sym}
\Gamma_i^T =
\left\{
\begin{array}{cc}
\Gamma_i, & i\leq p \\
&\\
-\Gamma_i, &  (p+1)\leq i\leq (p+q)
\end{array}
\right.
\end{equation}
can always assumed to be valid.

\setcounter{equation}0\section{An application: supersymmetries of
the free kinetic lagrangians of the `spinning' particle.}

We present for completeness the analysis of the extended
supersymmetric invariances for the simplest action of the
`spinning' particle model, given by the free kinetic term. We use
the quotation marks in the word `spinning' because actually the
considered actions describe particles with spin and are different
from the spinning particle models in which both fermions and
bosons are space-time vectors and bosons, in addition, are scalars
with respect to the supersymmetry transformations.

In general the
most significant dynamical systems are $\sigma$-models presenting
a  non-linear kinetic term; for such systems the extended
supersymmetries put constraints on the metric of the target. We
avoid entering this problem here and just limit ourselves to
illustrate how invariances under pseudo-Euclidean supersymmetries
can arise. We show in fact that a `spinning' particle evolving in a
non-Euclidean background in general admits invariances under
pseudo-Euclidean supersymmetries.
\par
We consider the models involving $d$ bosonic fields $X_a$ and $d$
spinors $\Theta_\alpha$ collected in the vector $\Psi$ \p{basis}
(no auxiliary fields are present).
\par
The free kinetic
action is given by
\begin{eqnarray}\label{Action}
S_K=\int dt{\cal L} &=& \frac{1}{2}\int dt\Psi^T\Lambda\Psi=
 \frac{1}{2}\int dt(X,\Theta)
\left(
\begin{array}{cc}
\lambda_1 H^2 & 0 \\
0& \lambda_2 H
\end{array}
\right)
\left(
\begin{array}{c}
X \\
\Theta
\end{array}
\right)\\
&=& \frac{1}{2}\int dt
({\dot X}_a\lambda_1^{ab}{\dot X}_b - i
\Theta_\alpha\lambda_2^{\alpha\beta}{\dot\Theta}_\beta )\label{lagr}
\end{eqnarray}
where the structure of the matrix $\Lambda$ is dictated by the
conservation of the fermion number and by dimensional arguments.
Both $\lambda_1, \lambda_2$ should be symmetrical in addition:
$\lambda_M^T=\lambda_M$.

The invariance of
the action under the supersymmetry transformations \p{genlaw1}
\begin{equation}\label{var}
  \delta{S_K}=\frac{i}{\sqrt{2}}\varepsilon^iX_a(\lambda_1\sigma_i-
\tilde\sigma^T_i\lambda_2)^{a\alpha} H^2 \Theta_\alpha =0,
\end{equation}
means that the following property of $\lambda$'s
\begin{equation}\label{Property}
  \lambda_1\sigma_i-
\tilde\sigma^T_i\lambda_2=0
\end{equation}
should be valid, in accordance with \p{Sym}
\begin{equation}\label{Sym1}
 \tilde\sigma^T_i=\eta^{ii}\sigma_i.
\end{equation}
It means that in the case of euclidean supersymmetry $(q=0)$ we
get
\begin{equation}\label{Identity}
  \lambda_1=\lambda_2=I,
\end{equation}
where $I$ is a $d$ dimensional identity matrix (see also
\cite{GR}). In the general case $(q\geq 1)$ the following
representation for the $\Gamma_i$ matrices \cite{O} is useful
\begin{equation}\label{Rep}
  \Gamma_\mu=\left(
\begin{array}{cc}
0 & \gamma_\mu \\
\gamma_\mu & 0
\end{array}
\right),\quad \mu=1,2\cdots,p+q-1,\quad
\Gamma_{p+q}=\left(
\begin{array}{cc}
0 & I \\
-I & 0
\end{array}
\right),
\end{equation}
where $\gamma_\mu$ form a real valued representation of the
Clifford algebra $C_{p,q-1}$ with the symmetry property \p{Sym}.
So, the conditions \p{Property} give, in particular,
$\lambda_1=-\lambda_2\equiv C$ and
\begin{equation}\label{Charge}
  C\gamma_\mu+\gamma_\mu^T C=0,
\end{equation}
which means that the matrix $C$ is the charge conjugation matrix
for the Clifford algebra $C_{p,q-1}$. The additional property of
symmetry for this matrix $C^T=C$ limits the possible signatures
$(p,q)$ for which the free action \p{Action} is invariant under
all $p+q$ supersymmetries. These possible signatures $(p,q)$ can
be represented by the following table {\scriptsize
\begin{center}
\begin{equation}\label{Total}
\begin{tabular}{|c|c|c|c|c|c|c|c|c|c|}
\hline
$p\setminus$ q&0&1&2&3&4&5&6&7&8
\\\hline
0&+&+&+&+&+&+&+&+&+
\\\hline
1&+&&+&&+&&+&+&+
\\\hline
2&+&&&+&+&+&+&+&+
\\\hline
3&+&&&&+&+&+&&+
\\\hline
4&+&+&+&+&+&+&+&+&+
\\\hline
5&+&&+&+&+&&+&&+
\\\hline
6&+&+&+&+&+&&&+&+
\\\hline
7&+&+&+&&+&&&&+
\\\hline
8&+&+&+&+&+&+&+&+&+
\\\hline
  \end{tabular}
\end{equation}
\end{center}
}which together with the modulo-$8$ periodicity gives the total answer.
For the cases of empty entries of the table it should be checked therefore
separately for each specific
choice of the matrices $\lambda_1, \lambda_2$ which
supersymmetries survive as invariances of the action.\par The
first non-trivial example concerns a $2$-dimensional `spinning'
particle ($d=2$). Its two bosonic and two fermionic degrees of
freedom carry the ${\bf \{2,2\}}$ representation of $(2,2)$
extended supersymmetry.
However, due to the condition (\ref{Property}) only half of these
supersymmetries can be invariances of the action. We obtain in fact
invariance under either the $(2,0)$ or the $(1,1)$ extended
supersymmetries, whether the target space is respectively
Euclidean or Minkowskian. Therefore already for the
$2$-dimensional Minkowskian `spinning' particle we observe the
arising of the pseudo-Euclidean supersymmetry invariance.

More generally, in all the cases except the euclidean one
$(q=0\quad or\quad p=0)$, exactly half of the eigenvalues of the
charge conjugation matrix $C$ are negative. It means that the
action \p{Action} describes the free motion in the spacetime with
signature $(d/2, d/2)$ with equal numbers of spacelike and
timelike coordinates. Both of them transform as irreducible
spinors of the isomorphisms group $SO(p,q)$ generated by
\begin{equation}\label{SOpq}
  J_{ik}=\frac{1}{4}\left[ \Gamma_i,\Gamma_k\right].
\end{equation}

 If one wants to have another
spacetime signature, some of the bosonic coordinates can be converted
into the auxiliary ones with the help of the procedure described at
the end of the previous Section. Formally it means that in the action
\p{Action} the time derivatives of some bosonic coordinates ${\dot X}_a$
are replaced by new auxiliary variables $F_a$.

The resulting representation ${\bf \{D,d,d-D\}}$ can, for example,
contain only $1$ timelike and $D-1$ spacelike bosonic dynamical
coordinates. Its corresponding action describes the `spinning'
particle with all its spacetime coordinates belonging to {\sl one}
irreducible representation of the extended supersymmetry. All the
additional $d-D$ bosonic coordinates are auxiliary. The example of
such description of the $4$ dimensional spinning particle with
$(4,4)$ extended supersymmetry was given in \cite{P}.

\setcounter{equation}0\section{Conclusions}

In this paper we presented some results concerning the
representation theory for irreducible multiplets of the
one-dimensional $N=(p,q)$ -- extended supersymmetry. As pointed
out in the text, a peculiar feature of the one-dimensional
supersymmetric algebras consists in the fact that the
supermultiplets formed by $d$ bosonic and $d$ fermionic degrees of
freedom accommodated in a chain with $M+1$ $(M\geq 2)$ different
spin states such as (\ref{rep}) uniquely determines a $2$-chain
multiplet of the form ${\bf \{d,d\}}$ which carries a
representation of the $N$ extended supersymmetry. Furthermore, it
is shown that all such $2$-chain irreducible multiplets of the
$(p,q)$ extended supersymmetry are fully classified; when e.g. the
condition $p-q=0$ mod $8$ is satisfied, their classification is
equivalent to those of Majorana-Weyl spinors in any given
space-time, the number $p+q$ of extended supersymmetries being
associated to the dimensionality $D$ of the spacetime, while the
$2d$ supermultiplet dimensionality is  the dimensionality of the
corresponding $\Gamma$ matrices. The more general case for
arbitrary values of $p$ and $q$ has also been fully discussed.
\par
These mathematical properties can find a lot of interesting
applications in connection with the construction of Supersymmetric
and Superconformal Quantum Mechanical Models. These theories are
vastly studied due to their relevance in many different physical
domains. To name just a few we mention the low-energy effective
dynamics of black-hole models, the dimensional reduction of
higher-dimensional superfield theories, which are a laboratory for
the investigation of the  spontaneous breaking of the
supersymmetry (for such investigations the extended supersymmetry
is an essential ingredient), as well as many others. As recalled
in the introduction, it is very crucial to build extended
supersymmetric models realized with the lowest-dimensional
representations. \par Another area in which we have started
applying the tools here elaborated is that one of supersymmetric
integrable hierarchies in $1+1$ dimensions. They are globally
supersymmetric non-linear non-relativistic theories, the
one-dimensional susies being realized through charges obtained by
integrating the supercurrents along the spatial line.
\vspace{.5cm}\\ \noindent{{\bf Acknowledgments}} We are grateful
to E.A. Ivanov and S.O. Krivonos for useful discussions and for
bringing to our attention the papers \cite{GR}. We are also
grateful to V. Zima for helpful remarks. F.T. ackowledges the
Bogoliubov Laboratory of Theoretical Physics, JINR, for the kind
hospitality. The work of A.P. was supported in part by the Russian
Foundation of Fundamental Research, under the grant 99-02-18417
and the joint grant RFFR-DFG 99-02-04022.

\end{document}